# Insight into the origin of Lithium/Nickel ions exchange in layered Li(Ni$_x$Mn$_y$Co$_z$)O$_2$ cathode materials

*Yinguo Xiao[a],\*, Tongchao Liu[a], Jiajie Liu[a], Lunhua He[b,d],\*, Jie Chen[c,d], Junrong Zhang[c,d], Ping Luo[c,d], Huaile Lu[c,d], Rui Wang[a], Weiming Zhu[a], Zongxiang Hu[a], Gaofeng Teng[a], Chao Xin[a], Jiaxin Zheng[a], Tianjiao Liang[c,d], Fangwei Wang[b,d], Yuanbo Chen[c,d], Qingzhen Huang[e], Feng Pan[a],\* and Hesheng Chen[c,d]*

[a] School of Advanced Materials Peking University Shenzhen Graduate School, Shenzhen, 518055, China
[b] Institute of Physics, Chinese Academy of Sciences, Beijing 100190, China
[c] Institute of High Energy Physics, Chinese Academy of Sciences, Beijing 100049, China
[d] Dongguan Neutron Science Center, Dongguan 523803, China
[e] NIST Center for Neutron Research, National Institute of Standards and Technology, 100 Bureau Drive, Gaithersburg, Maryland 20899, United States

*Corresponding authors
E–mail addresses: xiaoyg@pkusz.edu.cn (Y. Xiao), lhhe@iphy.ac.cn (L. He), panfeng@pkusz.edu.cn (F. Pan).

**Abstract**: In layered LiNi$_x$Mn$_y$Co$_z$O$_2$ cathode material for lithium-ion batteries, the spins of transition metal (TM) ions construct a two-dimensional triangular networks, which can be considered as a simple case of geometrical frustration. By performing neutron powder diffraction experiments and magnetization measurements, we find that long-range magnetic order cannot be established in LiNi$_x$Mn$_y$Co$_z$O$_2$ even at low temperature of 3 K. Remarkably, the frustration parameters of these compounds are estimated to be larger than 30, indicating the existence of strongly frustrated magnetic interactions between spins of TM ions. As frustration will inevitably give rise to lattice instability, the formation of Li/Ni exchange in LiNi$_x$Mn$_y$Co$_z$O$_2$ will help to partially relieve the degeneracy of the frustrated magnetic lattice by forming a stable antiferromagnetic state in hexagonal sublattice with nonmagnetic ions located in centers of the hexagons. Moreover, Li/Ni exchange will introduce 180° superexchange interaction, which further relieves the magnetic frustration through bringing in new exchange paths. Thus, the variation of Li/Ni exchange ratio vs. TM mole fraction in $LiNi_xMn_yCo_zO_2$ with different compositions can be well understood and predicted in terms of magnetic frustration and superexchange interactions. This provides a unique viewpoint to study the Li/Ni ions exchange in layered $Li(Ni_xMn_yCo_z)O_2$ cathode materials.

1. Introduction

The research and development of lithium ion batteries (LIBs) is indispensable for the application in not only mobile electronic devices but also large-scale devices, such as electric vehicles[1][2][3][4][5]. Many of current research activities to explore LIBs have focused on developing cathode materials that possess high energy density, high reversible capacity and stability. Among the well-known cathode materials, the layered transition metal oxides (TMO) $LiNi_xMn_yCo_zO_2$ (NMC, x+y+z = 1) is considered as a promising candidate owing to its high energy density (> 200 mA h $g^{-1}$)[6][7][8][9]. The $Li^+/Ni^{2+}$ exchange usually happens in NMC materials, referring to an antisite defect in which a $Li^+$ resides on $Ni^{2+}$ site (3b) and vice versa, a $Ni^{2+}$ occupies $Li^+$ site (3a). Obviously, the Li/Ni exchange will change the local structure of Li ion and consequently affect the electrochemical performance of NMC as cathode materials. For instance, it is argued that the Li/Ni exchange is detrimental to Li-ion diffusion due to narrowed layer space[10][11]. Moreover, the existence of Li/Ni exchange might cause damage to the thermal stability and induce strong distortion force, which could result in irreversible crystal structure transformation and the degradation of rate performance[12][13]. On the other side, some degree of cation disorder is considered as advantageous factors since it can mitigate the slab-distance concentration at high states of charge and deliver beneficial materials properties upon charge/discharge[14].

Insight into the variation of Li/Ni exchange ratio vs. different TMs mole fraction is important for tuning the performance of NMC cathode materials. Few different experimental approaches have been applied to probe antisite defect in layered $LiNi_xMn_yCo_zO_2$, such as x-ray powder diffraction and scanning transmission electron microscope[11][12][15][16]. However, it is nontrivial to obtain precisely structural information of Li-contained cathode material because the accuracy of measurement is largely dependent on the sensitivity of probes. Since a lithium atom has only three electrons, the cross sections of x-ray and electron scattering on Li atom are relatively small. In contrast, neutron diffraction is the method of choice to determine accurately the crystal and magnetic state of Li-contained material due to its high sensitivity to nuclei of lithium and magnetic moment as well. Furthermore, neutron diffraction also possesses the ability to tell the differences between Ni, Mn and Co since the neutron scattering lengths of these transition metals are very different. In order to gain a deep understanding of the origin and the effect of Li/Ni exchange in $LiNi_xMn_yCo_zO_2$ materials with various compositions, neutron powder diffraction experiments at different temperatures on $LiNi_xMn_yCo_zO_2$ cathode materials are imperative.

As a common feature in structural lattice, the Li/Ni exchange was previously reported in many $LiNi_xMn_yCo_zO_2$ compounds and it varies depending mainly on the mole fraction of TMs[11][17][18]. In NMC, the valence states of TM ions are also complicated because different compositions, charge distribution and electronegativities will result in the appearance of various valence states for TMs, e.g. $Ni^{2+}$, $Ni^{3+}$, $Co^{3+}$, $Co^{4+}$, $Mn^{4+}$ etc. Beside of imperfection in crystal lattice, another prominent structural features in NMC is that the TM spins construct a two-dimensional triangular networks, which can be considered as a simple case of geometrical frustration. As early as 1950, Wannier has studied the antiferromagnetism in triangular spin net theoretically[19]. It was found that the system with Ising spins on a triangular lattice is geometrically frustrated and hardly ordered at low temperatures. Because it is difficult for a geometrically frustrated system to construct a unique ground state with all antiferromagnetic interactions fully satisfied, the high ground-state degeneracy might take place in layered NMC and the long-range ordering of transition metal spins might be suppressed accordingly due to strong magnetic frustration. Combining with low temperature technique, neutron diffraction method can be adopted to probe the long-range magnetic order in NMC down to very low temperature. Indeed, the remaining fluctuations of transition metal spins can also be revealed by neutron diffraction results even if the long-range ordering of TM spins are significantly suppressed[20]. Moreover, the magnetic frustration might lead to nontrivial phenomena through the coupling with other degree of freedom for the relief of magnetic frustration[21][22]. Therefore, the results from neutron diffraction experiments hold the key to the elucidation of not only the origin of Li/Ni exchange but also the relationship between the Li/Ni exchange and magnetic frustration in layered NMC cathode materials. Naturally, the magnetic exchange interactions between spins of magnetic ions are of frustrated nature in magnetically frustrated spin systems.

In this work, we show that all investigated $LiNi_xMn_yCo_zO_2$ cathode materials with different proportion of transition metal ions exhibit antisite defects, i.e. Li/Ni exchange or cation disorder, in crystal lattice instead of a perfectly ordered crystal lattice. The Li/Ni exchange ratio varies from 1.6(2)% to 6.3(2)% in seven different $LiNi_xMn_yCo_zO_2$ materials depending on the concentration of three individual transition metals. By comparison, it was found that increasing Ni and/or Mn contents would promote the formation of Li/Ni exchange, whereas Co has exactly the reverse effect. Given that the magnetic frustration in $LiNi_xMn_yCo_zO_2$ is inherent as revealed by combined low temperature neutron diffraction and magnetization measurements, it is speculated that the Li/Ni exchange formed in $LiNi_xMn_yCo_zO_2$ is an inevitable way to partially lift spin degeneracy and to relieve magnetic frustration in triangle lattice of TM layers.

## 2. Experimental Section

**Materials Synthesis**: The coprecipitation method was employed to synthesized spherical $[Ni_xMn_yCo_z](OH)_2$ (x=1/3, 0.4, 0.42, 0.5, 0.6, 0.7, 0.8) precursors of NMC. The appropriate amounts of $NiSO_4 \cdot 6H_2O$, $CoSO_4 \cdot 7H_2O$, and $MnSO_4 \cdot 5H_2O$ were dissolved in water to prepare an aqueous solution with a concentration of 2.0 mol $L^{-1}$, then the mixed solution was added into a continuously stirred tank reactor (CSTR, 4 L) under a $N_2$ atmosphere as the starting materials for the synthesis of $[Ni_xMn_yCo_z](OH)_2$ (x=1/3, 0.4, 0.42, 0.5, 0.6, 0.7, 0.8). Simultaneously, a 4.0 mol $L^{-1}$ NaOH solution (aq.) and the desired amount of NH4OH solution (aq.) as a chelating agent were separately pumped into the reactor. Finally, the concentration of the precursor solution is 0.5 mol $L^{-1}$, the pH value of the precursor solution is kept at 10, the temperature is kept at 60 °C, and the stirring speed is kept at 400 rpm/s. The precursor powders were obtained through filtering, washing, and drying in a vacuum oven overnight. The precursors $[Ni_xMn_yCo_z](OH)_2$ (x=1/3, 0.4, 0.42, 0.5, 0.6, 0.7, 0.8) and LiOH $H_2O$ were mixed by 1:1.03 molar ratio, then were transferred into the tube furnace. The sintering procedures were done at 825-900 °C for different components of the precursors to achieve good ordered structures: 900 °C for $Li(Ni_{1/3}Mn_{1/3}Co_{1/3})O_2$, 900 °C for $Li(Ni_{0.4}Mn_{0.4}Co_{0.2})O_2$, 900 °C for $Li(Ni_{0.42}Mn_{0.42}Co_{0.16})O_2$,, 850 °C for $Li(Ni_{0.5}Mn_{0.3}Co_{0.2})O_2$, 825 °C for $Li(Ni_{0.6}Mn_{0.2}Co_{0.2})O_2$, 825 °C for $Li(Ni_{0.7}Mn_{0.15}Co_{0.15})O_2$, and 825 °C for $Li(Ni_{0.8}Mn_{0.1}Co_{0.1})O_2$. It should be noted that the sintering temperatures are already optimized to get the least Li/Ni exchange ratio for each compounds and it will be presented and discussed in another work.

**Structural and Magnetic Properties Characterization**:

Neutron powder-diffraction measurements were performed on two neutron diffractometers, i.e. high resolution powder diffractometer BT1 at National Institute of Standards and Technology (NIST), USA, and time-of-flight (TOF) diffractometer GPPD (General Purpose Powder Diffractometer) at China Spallation Neutron Source (CSNS), Dongguan, China. At BT1, Cu(311) and Ge(311) monochromator were used to produce a monochromatic neutron beam of wavelength 1.5397 Å and 2.0775 Å, respectively. The samples were loaded in a vanadium sample holder and then installed in the liquid helium cryostat that can generate temperature down to 3 K. The neutron powder diffraction data were collected at both 300 and 3 K. At GPPD, samples were loaded in 9.1 mm diameter vanadium cans and neutron diffraction patterns were collected at room temperature with wavelength band from 0.1 to 4.9 Å. It is worth noting that the CSNS is China's first pulsed neutron source and GPPD is the first TOF neutron powder diffractometer at CSNS[23][24], which has been constructed very recently. The program FULLPROF[25] was used for the Rietveld refinement of the crystal structures of the compounds. A Quantum Design Physical Property Measurement System (PPMS) was used to characterize the magnetic properties of $LiNi_xMn_yCo_zO_2$ powder samples in the temperature range from 2 to 390 K and the magnetic field range from 0 to 5 T. The temperature dependence of magnetization was measured during warming from 2 to 390 K under an applied magnetic field of 1000 Oe.

## 3. Results and Discussion

The crystal structure of $LiNi_xMn_yCo_zO_2$ can be described as the alternate stacking of Li layers and the layers composed of edge-sharing $TMO_6$ octahedra as illustrated in Figure 1(a), thus the crystal electric field generated from six surrounding oxygen ligands will act on the TM ion and remove the energetic degeneracy of the 3$d$ orbitals of TM. Consequently, the splitting of the $d$-orbitals of TM ions will occur and give rise to two orbital sets, i.e. doubly degenerate $e_g$ orbitals and triply degenerate $t_{2g}$ orbitals (Figure 1(b)). The $e_g$-electrons with higher energy generally lie close to the Fermi level and have a relatively strong hybridization with neighboring oxygen. Although $t_{2g}$-electrons are more localized, they can have a small overlap with the oxygen 2*p*-states, which will also lead to the superexchange interaction. The sign and the magnitude of superexchange interaction between TM ions largely depend on the orbital character of the localized charge, which can be explained and described in terms of Goodenough-Kanamori-Anderson (GKA) rules[26][27][28]. Very recently, theoretical calculations by Chen *et al.* and experimental measurements by Chernova *et al.* suggested the potential impact of superexchange interaction on the structural ordering, electrochemical property and thermal stability of $LiNiO_2$ and Co-, Mn-, and Al-substituted variations[29][30], indicating the possible correlation of some types of magnetic interactions with the structural and electrochemical properties in layered $LiTMO_2$ cathodes. Thus, superexchange interactions contain competing neighboring components are the dominant magnetic interactions in $LiNi_xMn_yCo_zO_2$ materials.

We choose seven representative $LiNi_xMn_yCo_zO_2$ compounds with different TM compositions for systematically experimental studies: $Li(Ni_{0.33}Mn_{0.33}Co_{0.33})O_2$ (333), $Li(Ni_{0.4}Mn_{0.4}Co_{0.2})O_2$ (442), $Li(Ni_{0.42}Mn_{0.42}Co_{0.16})O_2$ (424216), $Li(Ni_{0.5}Mn_{0.3}Co_{0.2})O_2$ (532), $Li(Ni_{0.6}Mn_{0.2}Co_{0.2})O_2$ (622), $Li(Ni_{0.7}Mn_{0.15}Co_{0.15})O_2$ (71515), $Li(Ni_{0.8}Mn_{0.1}Co_{0.1})O_2$ (811). Generally, magnetic materials containing 3*d* TM ions can achieve long-range magnetic order as magnetic ground state at low temperatures[31][32]. However, the magnetic ordering in magnetic material can be suppressed to very low temperature when the low energy ground state is frustrated, arising from geometrical constraints or competing exchange interactions[22][33]. In NMC, the TM spins form a 2D triangular lattice separated by Li layers. This is a typical ingredient for geometrical frustration, since it is not possible to minimize the interaction energy for all pairs of nearest-neighbor spins in magnetic triangular lattice. As shown in Figure 1(c), no matter how up and down spins are arranged on the triangle, there is always at least one uncertain spin in a triangle. It might point up or down once its neighbouring spins are arranged antiferromagnetically on a triangular lattice, indicating that at least one pair of parallel spins exists on a triangle although exchange interaction favors antiparallel neighboring spins. If we arrange spins in purpose so that only one bond connecting parallel spins on each triangle, then we can get a ground state, which is a degenerate manifold. The number of such ground states will increase exponentially with the increase of system size. However, it is in violation of the third law of thermodynamics and gives an finite entropy at absolute zero temperature[19]. Apparently, large ground state degeneracy will suppress the ordering of spins.

In order to confirm our speculation on the existence of strong magnetic frustration and approach the magnetic ground state in NMC compounds, we firstly carried out neutron powder diffraction measurements for (333) at low temperature. Among all NMC compounds, (333) is a simple example to demonstrate the structural and magnetic inhomogeneity as it contains heterovalent ions $Ni^{2+}$, $Co^{3+}$ and $Mn^{4+}$ with the same contents. Upon cooling down to 3 K, the rhombohedral structure of (333) persists, except the shift of nuclear reflections from low-Q to high-Q region, indicating the shrink of lattice constant as indicated in Figure 2(a). It is worthwhile to note that no extra reflection is observed besides nuclear ones, in other words, long-range magnetic order does not exist with temperature as low as 3 K. Similarly, long-range magnetic order is also absent in (442) sample [see Figure 2(b)]. Nevertheless, the magnetic diffuse scattering signal can be observed clearly in both (333) and (442) at 3 K as demonstrated in Figure 2(c) and (d). The appearance of diffuse scattering implies the existence of short-range antiferromagnetic spin correlation and pronounced magnetic frustration, as symbolically illustrated in Figure 2(e). It is also noticed that the broad diffuse peaks are slightly asymmetric, which is characteristic of a two dimensional short-range order[34] and agrees well with the two dimensional character of NMC materials.

The absent of long-range magnetic order is further confirmed by the magnetization measurements of (333) and (442) as a function of temperature. Figure 3 (a) and (b) shows the temperature dependence of the magnetization and inverse susceptibility of (333) and (442) measured under magnetic field of 0.1 T, respectively. The temperature dependence of the magnetization in these compounds does not exhibit any anomaly instead of a steady and monotonic increase in magnetization with decreasing temperature, suggesting the dominant antiferromagnetic interaction. It is known that the magnetic susceptibility of localized noninteracting magnetic ions in high temperature region can be written as: $\chi(T) = \frac{C}{T - \Theta_{CW}}$, where $C$ and $\Theta_{CW}$ are the Curie constant and Curie-Weiss temperature, respectively. As indicated in Figure 3 (b), the inverse susceptibilities of both (333) and (442) strictly follow the Curie-Weiss behavior in high temperature range as they can be fitted properly with a Curie-Weiss function, while the inverse susceptibility 1/χ deviated from the Curie-Weiss estimation below 180(4) K in 333 and 191(4) K in 442, indicating the onset of considerable magnetic correlations. The Curie-Weiss temperatures $\Theta_{CW}$ of (333) and (442) are deduced to be -106(1) and -111(1) K, respectively. The obtained Curie-Weiss temperature with negative value also hints that the coupling between TM ions are dominated by antiferromagnetical interactions. In general, the frustration parameter in terms of $f = |\Theta_{CW}|/T_N$ can be used to evaluate the strengthen of magnetic frustration in magnetic system[33]. Clearly, $f > 1$ corresponds to frustration and $f > 10$ are typically taken as empirical evidence of a highly frustrated magnet. Given the fact that the $|\Theta_{CW}|$s are higher than 100 and long-range magnetic order is prevented to below 3 K for both (333) and (442), the frustration parameters of these compounds are estimated to be larger than 30, indicating a strong spin frustrated system. Such strong frustration arise from the competition between local anisotropy and the strong conflicting interactions associated with the moments and structural character of the strongly geometrically frustrated magnets. According to the mean field approximation, it is assumed that each magnetic atom experiences a field proportional to the macroscopic magnetization. If only the nearest-neighbor interactions between TM ions are taken into account, the effective exchange parameter $J_{eff}$ is related to the Curie-Weiss temperature via $|J_{eff}|/k_B = \frac{3}{2} \cdot \frac{|\Theta_{CW}|}{S(S+1)z}$, where S is the spin quantum number of the TM ion, z = 6 is the number of nearest neighbors[35]. $|J_{eff}|$ is then determined to be around 0.61 meV for (333) if we adopt S = 3/2 as average spin quantum for $Ni^{2+}$ (S = 1), $Mn^{4+}$ (S = 3/2) and $Co^{3+}$ (S = 2) ions.

In spite of the intrinsic spin frustration, the magnetic frustrated system will always find a peculiar way to lift its degeneracy by changing the crystal symmetry accompanied with the generation of non-collinear spin ordering or orbital ordering[36][37][38]. As illustrated in Figure 1(c), the replacement of magnetic $Ni^{2+}$ ion by nonmagnetic $Li^+$ ion in TM layer of $LiNi_xMn_yCo_zO_2$ structure will break the local crystal symmetry, change the exchange interaction, and stabilize the spin state of surrounding TM spins in an antiferromagnetical arrangement on a honeycomb lattice (Situation I in Figure 1(c)). Consequently, the frustration is relieved accordingly in local region. It is noted that $Co^{3+}$ is in low spin state in NMC and of nonmagnetic because of the absence of unpaired electron in the electronic configuration. Therefore, the appearance of $Co^{3+}$ in the triangular lattice hold the similar effect on relief of magnetic frustration as that of nonmagnetic $Li^+$ ions (Situation II in Figure 1(c)). Obviously, (333) contains the largest amount of $Co^{3+}$ among all investigated $LiNi_xMn_yCo_zO_2$ compounds. Since nonmagnetic $Co^{3+}$ can relieve magnetic frustration, the strength of magnetic frustration in (333) should be the smallest compared with other compounds. Thus, less Li/Ni exchange ratio is needed to lift the degeneracy of magnetic ground state. On the contrary, more Li/Ni exchange need to be introduced into lattice in order to relieve magnetic frustration in NMC with large magnetic moments in TM layers, such as (424216) with large amount of magnetic $Ni^{2+}$ and $Mn^{4+}$ ions as well as (811) with considerable amount of magnetic $Ni^{2+}$ and $Ni^{3+}$ ions.

In order to validate above speculation that Li/Ni exchange is related to the strength of magnetic frustration, we performed neutron powder diffraction experiments on a series of NMC materials to determine their Li/Ni exchange ratio precisely. As illustrated in Figure 4, all $LiNi_xMn_yCo_zO_2$ compounds are isostructural and crystallized in rhombohedral structure with space group *R-3m* at 300 K. The experimental neutron powder diffraction pattern of (333) at 300 K is shown in Figure 4(a), together with Bragg position, the calculated pattern, and the differences between the experimental and the calculated patterns. For the Rietveld refinement of crystal structure, we first adopted the disordered arrangement model for all three individual TM, i.e. with Li partially occupies 3*b* site whereas the same amount of TM (either Ni or Mn or Co) occupies 3*a*, for the structural refinement. It was found that the smallest reliability factor of the refinement can only be obtained with considering the Li/Ni exchange model, which confirmed the Li/Ni exchange scheme instead of Li/Mn or Li/Co exchange. The Li/Ni exchange ratio in (333) is determined to be 1.6(2)%. Assuming that the thermal motions of atoms is isotropic, the thermal parameter $B_{iso}$ of Li(3a), Ni/Mn/Co(3b) and O(6c) sites are deduced to be 0.87(8), 0.24(5) and 0.61(2) $Å^2$, respectively. The thermal parameter relates to the atomic thermal motion in crystal lattice and reflects the presence of static atomic disorder.

Subsequently, same procedures for structural refinement are also applied to analyse the neutron powder diffraction data of other compounds, as shown in Figure 4(b)-(h). The best fit of neutron powder diffraction patterns for (333), (622), (532), (71515), (442), (811) and (424216) are obtained with the lattice parameters and Li/Ni exchange ratio summarized in Table 1. The Rietveld refinement results of (333), (622), (532), (71515) and (442) are based on the data collected at BT1 (Figure 4(a)-(e)), while the refinement results of (811), (424216) and (442) are based on the data collected at GPPD (Figure 4(f)-(h)). Because a wide wavelength band from 0.1 to 4.9 Å is adopted for the measurements at GPPD and the neutron absorption cross section is inversely proportional to the velocity of neutron (so-called "1/*v* law"), the background of neutron diffraction pattern from GPPD increases gradually with increasing Q, which is in contrast to the flat background obtained from BT1 with constant neutron wavelength. The (442) sample is chosen as proof sample for measurements on both BT1 (Figure 4(e)) and GPPD (Figure 4(h)). The structural information of (442), especially the Li/Ni exchange ratio, deduced from these two instruments exhibit excellent agreement, although the lattice parameters are slightly different probably due to the temperature variation. The detailed results of refinement including atomic positions, thermal parameters and reliability factors for all investigated $LiNi_xMn_yCo_zO_2$ compounds are listed in Table S1 and S2 of the Supplemental Information.

The Li/Ni exchange ratio in isostructural NMC compounds varies from 1.6(2)% to 6.3(2)%, depending on the proportion of TMs. To enlighten the influence of individual TM ion on Li/Ni exchange ratio, the Li/Ni exchange for each compounds are plotted in ternary phase diagram of $LiNiO_2$-$LiMnO_2$-$LiCoO_2$ system, as shown in Figure (i) and (j). It is clear that the lowest Li/Ni exchange ratio is observed in (333), which possesses the largest amount of Co. The observations are in good agreement with previous reports, in which the Li/Ni exchange is found to more likely take place in $LiNi_xMn_yCo_zO_2$ compounds with less concentration of Co ions[39][40]. Greater Li/Ni exchange are observed for compounds with rich Ni and Mn, such as in (424216) and (811). It is also noticed that the variation of Li/Ni exchange ratio with composition is nonlinear. For instance, the Li/Ni exchange ratio of $LiNi_xMn_yCo_zO_2$ (with x = y and x+y+z = 1) increases from 1.6(2) to 4.2(2) with x increase from 0.33 to 0.4, while it increases from 4.2(2) to 6.3(2) with x increase only from 0.4 to 0.42. As discussed previously, nonmagnetic $Co^{3+}$ can act as an ingredient to relieve magnetic frustration by forming a nonmagnetic center on a honeycomb lattice unit so that the surrounding TM spins can be stabilized in an antiferromagnetic arrangement. Because (333) contains the largest amounts of $Co^{3+}$ among all NMC compounds, the strength of magnetic frustration in (333) is expected to be the smallest compared with others. In addition, the net magnetization at 2 K are obtained to be 0.5 emu/g for (333), which is half of the net magnetization of (442) obtained under the same magnetic field of 1000 Oe as indicated in Figure 3(a), reflecting less frustration in (333). According to our previous works[41][42], the formation of Li/Ni exchange in NMC compounds was understood from the viewpoint of formation energy for one pair of Li/Ni exchange, which is associated with the strengthen of superexchange interaction between TM cations via bridged O anion. This also supports our present arguments, since magnetic frustration is augmented by disorder under the scheme of superexchange interactions in NMC compounds. In other words, the underneath mechanism of the Li/Ni exchange is not only associated with the superexchange interaction but also related essentially to the magnetic frustration in NMC compounds. Generally, isotropic short-range Heisenberg interactions are present in a real magnetic system. The intralayer exchange interaction $J1$ between TM ions in NMC dominates in layered structure if there is no Li/Ni exchange. The occurrence of Li/Ni exchange will lead to new superexchange pathways between disordered $Ni^{2+}$ located in $Li^+$ slab and neighbouring TM ions remain in TM slabs (Situation III in Figure 1(c)). The new superexchanges are labeled as $J2$ and $J3$ in Figure 1(c). It is also noticed that the neighboring numbers of both $J2$ and $J3$ are six, indicating the considerable weight of exchange interaction between $Ni^{2+}$ in Li slab and neighboring TM ions in TM slab. Because the new exchange interaction pathways are out-of-plane, the replacement of $Ni^{2+}$ at $Li^+$ site can also relieve the magnetic frustration. In contrast to the 90° intralayer exchange interactions, the spins of $Ni^{2+}$ at $Li^+$ site are coupled antiferromagneticaly to TM spins in the TM layer by strong 180° exchange. Among all 180° superexchange interactions, the $Ni^{2+}$—$O^{2+}$—$Ni^{2+}$ is the strongest, followed by $Ni^{2+}$—$O^{2+}$—$Mn^{4+}$ interaction with strong strength, whereas the $Ni^{2+}$—$O^{2+}$—$Co^{3+}$ is very weak [41]. Therefore, increasing $Co^{3+}$ will lead to the suppression of Li/Ni exchange, reflected from the increase in formation energy. It is worthwhile to note that the calculated formation energy of Li/Ni anti-site defect can only exhibit reliability when the spin polarization of Ni, Co and Mn ions is taken into account, indicating the important role of superexchange interaction in Li/Ni disordering in NMC compounds [41].

Combining neutron diffraction results and our speculation concerning the relationship between Li/Ni exchange ratio and strengthen of magnetic frustration, it can be proposed that both relief of magnetic frustration and the interlayer superexchange interaction by Li/Ni exchange in layer NMC compounds can reduce the system energy and enhance the structure stability. The insight of the variation of Li/Ni exchange ratio vs. different TMs mole fractions (Ni/Mn/Co) in Li(Ni$_x$Mn$_y$Co$_z$)O$_2$ can be revealed, for example, NMC(333) with more Co$^{3+}$ (nonmagnetic with low-spin $d$6 configuration as shown in Figure 1(b)) generates less magnetic frustration to lead to lower Li/Ni exchange ratio compared with other NMCs. In contrast, (424216), which contains less Co$^{3+}$ but more Mn$^{4+}$ and Ni$^{2+}$ (both are strong magnetic with S = 3/2 and S = 1, respectively) generates stronger magnetic frustration than other MNCs, so that larger Li/Ni exchange ratio has to be occurred in order to relieve frustration and to create strong the Ni$^{2+}$—O$^{2+}$—Ni$^{2+}$ and Ni$^{2+}$—O$^{2+}$—Mn$^{4+}$ 180° superexchange interactions for enhancing structure stability.

## 4. Conclusion

In summary, we present results of a systematic study on the Li/Ni exchange in layered Li(Ni$_x$Mn$_y$Co$_z$)O$_2$ cathode material through experimental approach with neutron diffraction and magnetization measurements. The Li/Ni exchange ratios in series of LiNi$_x$Mn$_y$Co$_z$O$_2$ materials have been precisely determined for the first time through high resolution neutron diffraction experiments by using BT1 at NIST and GPPD at CSNS. Combining the magnetization and neutron diffraction results, it is evident that the magnetism of LiNi$_x$Mn$_y$Co$_z$O$_2$ is intrinsically frustrated. Nevertheless, the magnetic frustration can be partially relieved by introducing structural inhomogeneity via Li/Ni disordering. Based on the correlated relation between Li/Ni exchange ratio and strengthen of magnetic frustration, we argue that Li/Ni exchange can act as an effective way to relieve the frustration in $LiNi_xMn_yCo_zO_2$, furthermore, the variation of Li/Ni exchange ratio vs. different TMs-ratio of NMC can be explained and predicted. Our findings might shed light on understanding the mechanism of the formation of Li/Ni antisite defect in NMC and will help with the optimization of ingredients so that to promote the electrochemical performance of NMC materials.

**Acknowledgements**

This work was financially supported by National Materials Genome Project (2016YFB0700600 and 2016YFB0100106), the National Natural Science Foundation of China (Nos. 21603007 and 51672012), Shenzhen Science and Technology Research Grant (Nos. JCYJ20150729111733470 and JCYJ20151015162256516) and Youth Innovation Promotion Association CAS (Grant No. 2017023). The authors are grateful to Prof. Baogen Shen for valuable discussion and substantial support.

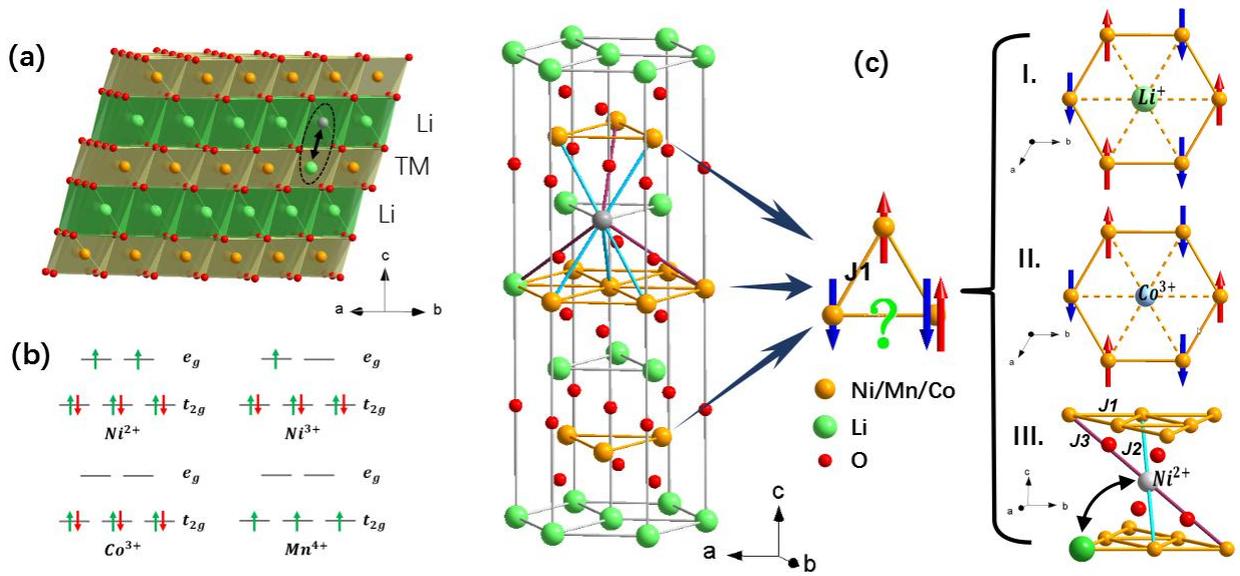

**Figure 1.** (a) Schematic drawing of the layered crystal structure of $LiNi_xMn_yCo_zO_2$ compounds. (b) Schematic diagram of crystal field splitting of various TM ions in $LiNi_xMn_yCo_zO_2$. (c) Illustration of a pair of Li/Ni exchange and arrangement of magnetic moments in the TM layer of triangular lattice. The local structure and exchange interactions are modified upon the Li/Ni exchange. Three different ways to relieve magnetic frustration are denoted as I, II, and III. The paths of exchange interactions are labeled as $J1$, $J2$ and $J3$.

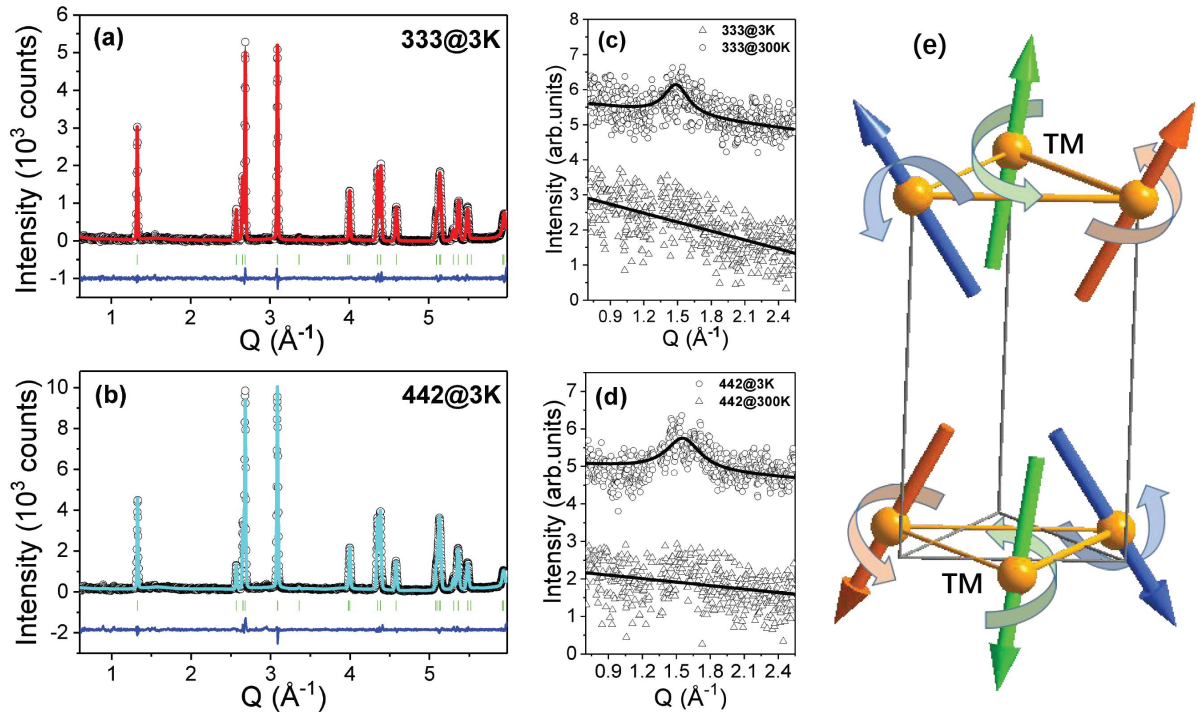

**Figure 2.** Neutron powder diffraction refinement patterns for (333) (a) and (442) (b) compounds. The circles represent the observed intensities, the solid line is the calculated pattern. The difference between the observed and calculated intensities is shown at the bottom. The vertical bars indicate the expected Bragg reflection positions. The magnetic diffuse scattering signals of (333) and (442) at 3 K are shown in (c) and (d), respectively, with the neutron data at sample Q-range as comparison. (e) Schematic view of antiferromagnetic spin fluctuation of transition metal spins in the triangular lattice of transition metal layer.

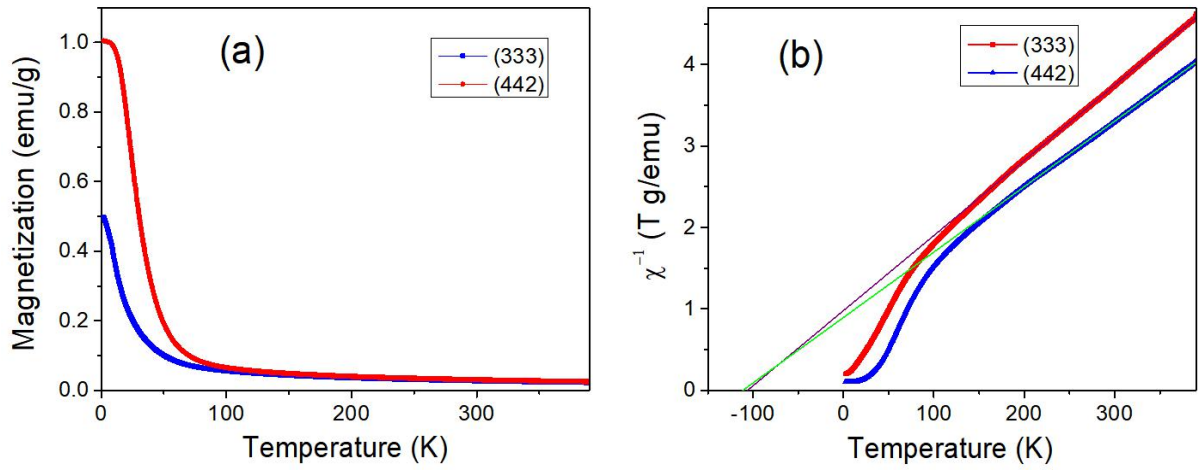

**Figure 3.** Temperature dependence of the magnetization (a) and inverse magnetic susceptibility (b) of (333) and (442) measured under magnetic field of 1000 Oe.

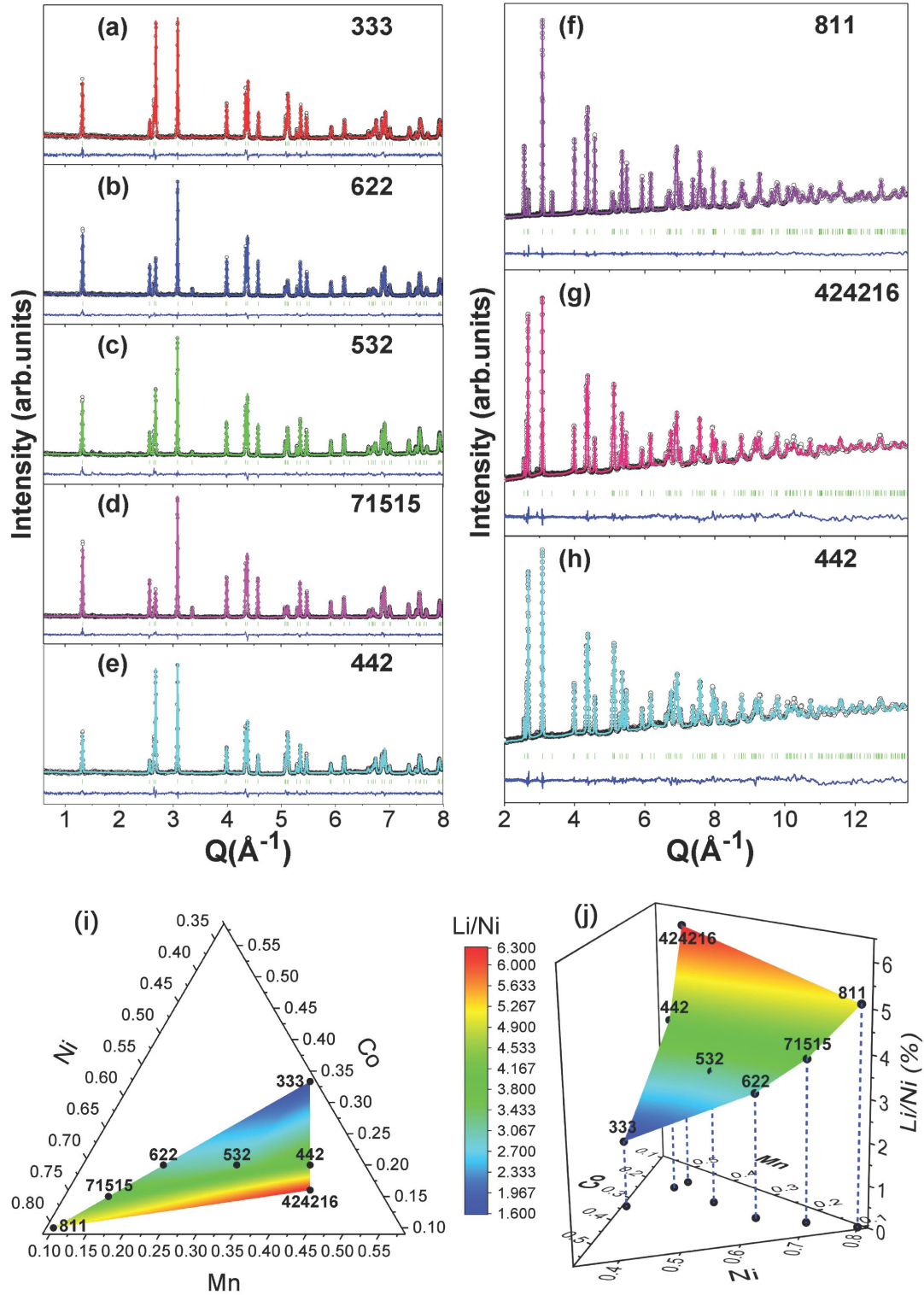

**Figure 4.** Neutron powder diffraction refinement patterns for series of $Li_{1-x}Ni_xMn_yCo_zO_2$ compounds based on the data collected at BT1(a)-(e) and GPPD(f)-(h), respectively. (i) and (j) Li/Ni exchange ratios of $Li_{1-x}Ni_xMn_yCo_zO_2$ compounds with different compositions of transition metals.

**Table 1.** Lattice parameters and Li/Ni exchange ratios $\delta$ of Li(Ni$_x$Mn$_y$Co$_z$)O$_2$ materials.

| Sample | a (Å) | b (Å) | c (Å) | V (Å$^3$) | $\delta$(%) |
|---|---|---|---|---|---|
| 333 | 2.8602(1) | 2.8602(1) | 14.2315(3) | 100.82(1) | 1.6(2) |
| 622 | 2.8688(1) | 2.8688(1) | 14.2206(3) | 101.35(1) | 2.9(2) |
| 532 | 2.8692(1) | 2.8692(1) | 14.2410(3) | 101.53(1) | 3.2(2) |
| 71515 | 2.8719(1) | 2.8719(1) | 14.2147(3) | 101.53(1) | 3.8(2) |
| 442 | 2.8720(1) | 2.8720(1) | 14.2603(3) | 101.86(1) | 4.2(2) |
| 811 | 2.8714(1) | 2.8714(1) | 14.1973(3) | 101.37(1) | 5.1(2) |
| 424216 | 2.8712(1) | 2.8712(1) | 14.2610(3) | 101.81(1) | 6.3(2) |